\documentclass[12pt,preprint]{aastex}

\slugcomment{}

\shorttitle{The environmental Dependence of Star Formation and AGN Activity}
\shortauthors{Reviglio, Helfand}

\begin{document}

\title{On radio-bright Active Galactic Nuclei in a complete Spectroscopic Redshift Survey}

\author{Pietro Reviglio,  David J. Helfand}
\affil{Astronomy Department, Columbia University, New York, NY 10027}

\email{reviglio@astro.columbia.edu, djh@astro.columbia.edu}

\begin{abstract}
Analysis of the frequency and physical properties of galaxies with  star-formation and AGN activity in different environments in the local universe is a cornerstone for understanding structure formation and galaxy evolution. 
We have built a new multiwavelength catalog for galaxies in a complete redshift survey (the 15R Survey), gathering information on their $H\alpha$, R-band, radio, far-infrared, and X-ray emission, as well as their radio and optical 
morphologies, and have developed a classification scheme to compare different selection methods and to select accurately samples of radio emitting galaxies with AGN and star-forming activity. While alternative classification schemes do not lead to major differences for star-forming galaxies, we show that spectroscopic and photometric classifications of AGN lead to incomplete samples.
In particular, a large population of AGN-containing galaxies with absorption-line spectra, and in many cases extended radio structures (jets, lobes), is missed in the standard Baldwin-Phillips-Terlevich emission-line classification of active galaxies. This missed class of objects accounts for roughly half of the radio AGN population. Similarly, for X-ray selected AGN in our sample, we find that absorption-line AGN account for half of the sample.
Spectroscopically unremarkable, passive galaxies with AGN activity are not an exception, but the norm, and we show that although they exist in all environments, these systems preferentially reside in higher density regions. Because of the existence of this population, the fractional abundance of AGN increases with increasing density, in contrast to the results based on
emission-line AGN extracted from the 15R, Sloan and 2DF redshift surveys. Since  emission-line radio AGN are mostly associated with late-type galaxies and absorption-line radio AGN with early-type galaxies, the trends found are connected to the well-known but poorly understood density-morphology relation. 

\end{abstract}


\section{Introduction}

Models of galaxy formation and evolution require a clear understanding of star-formation and AGN activity in galaxies and how these are related to other galaxy properties such as morphological type, gas and stellar content, and age. The link between the presence of these types of activity and the surrounding environment may also shed light on the process of structure formation and its evolution. Emission lines in galaxy spectra are indicative of processes that
excite and ionize neutral hydrogen and metals present in the interstellar 
medium. Two major processes
are known to lead to the formation of emission-line spectra:
star-formation, where the ionizing radiation is primarily the UV radiation of  hot, young, massive stars, and the presence of an active galactic nucleus, where the ionizing radiation is generally attributed to the process of accretion onto a supermassive black hole residing in the galaxy's center.

AGN in particular show a variety of features in their emission line spectra: Seyfert 1 galaxies have very broad emission lines that include both allowed and forbidden lines (HI, HeI, HeII, [OIII]) while Seyfert 2 galaxies have only narrow lines. Similarly, radio galaxies have been observed to divide into two families, one with broad emission lines and the other exhibiting only
narrow lines or even lacking emission lines entirely \cite[]{Antonucci93}

Baldwin, Phillips and Terlevich (BPT; 1981) first pointed out that, in general,
AGN and starforming galaxies can be distinguished on the basis of the ratio of their emission lines: AGN should have greater ratios of [OIII]5007 to the $H\beta$ line because the creation of [OIII] by photoionization demands photons above 35~eV which are rare in stellar spectra but common in AGN continua; similarly,
the [NII]6583/H$\alpha$ ratio should be higher in AGN. BPT showed that when these two quantities are plotted for emission-line galaxies, samples separate into two families of AGN-dominated and star-forming galaxies. The BTP method 
and various improvements thereto \cite[]{Veilleux87, Dopita2000} is now routinely used to classify emission-line galaxies.

One major use of this diagnostic is the classification of galaxies in large redshift surveys in order to understand the dependence of star-formation
and AGN activity on the physical properties of the host galaxy and the environment in which it resides. For spectroscopically selected samples, it has been established that the fraction of emission line galaxies decreases with increasing galaxian density, while the fraction of absorption-line galaxies shows an increasing trend in dense environments \cite[]{Carter01, Mateus04}.
When emission-line galaxies are split into two families using the BTP diagnostic, these studies have shown that the fractional abundance of star-forming galaxies decreases with increasing environmental density. The trend  for AGN is unclear, however. The fraction of emission-line-classified AGN does not vary significantly with density in several recent studies based on redshift surveys such as 15R  \cite[]{Carter01}, SDSS \cite[]{Miller03}, and 2DF \cite[]{Mateus04}. Kauffman et al. \cite[]{Kauffmann04} find that AGN with strong [OIII] line emission tend to reside preferentially in low-density environments.

An accurate constraint on the environmental dependence of AGN activity can shed light on the complex interelationship between environment, galaxy morphology, and activity, as well as help to define the properties of the 
mechanism(s) powering active galactic nuclei. Different models for AGN activity predict different fractions of AGN in different environments. If, for example, the fuel of AGN activity is the cold gas that also fuels star-formation, AGN should follow the decreasing trend with increasing density found for star-forming galaxies (e.g., Kauffmann et al. 2004)
If instead the critical factor is the supermassive black hole which is 
connected to the existence of a bulge, then the distribution of AGN should trace the distribution of bulges. Even less understood is the interplay between 
AGN activity and star-formation activity in a galaxy. Recent studies of the hosts of AGN have shown that powerful Seyfert 2 galaxies have young stellar populations \cite[]{Kauffmann03}, suggesting the possibility that strong AGN activity requires host galaxies rich in cold gas.

Radio emission at 1.4 GHz is a tracer of both star-formation and AGN activity, and has the great advantage of being unaffected by dust obscuration. It offers an alternative to optical line emission for pursuing studies of the
dependence of galactic activity on environment and other factors.
Moreover, radio morphologies provide an independent way to classify radio-emitting galaxies as predominantly star-forming galaxies or AGN. Star-forming galaxies show extended radio emission arising through the interaction of cosmic rays accelerated by  supernova explosions with interstellar magnetic fields \cite[]{Condon92}. AGN, on the contrary, have point-like emission or strong radio jets and lobes. Given the relatively
high sensitivity thresholds for radio surveys, only brighter star-forming galaxies and AGN are detectable. Nevertheless, these constitute a well-defined sub-sample of a larger optical galaxy catalog.

The analysis presented in this paper has been carried out on a sample of
radio-emitting galaxies extracted from  the 15R Redshift Survey \cite[]{Carter01}.
This survey has the advantage that the median redshift is sufficiently low
that the angular resolution of existing radio surveys provides a strong
morphological diagnostic; in addition, it is small enough to allow the visual 
classification of the morphologies for all selected radio sources and their host galaxies. These points are crucial for studies of radio-selected samples, since the complicated morphology of radio sources, jets in particular, renders problematic automated selection and classification algorithms.  The techniques developed in this work have been the basis for extending our work to larger samples of galaxies. The results obtained in this paper have been confirmed using two larger statistical samples drawn from the 2DF and the SDSS 2DR Surveys, the results of which will be presented in a subsequent paper. They
are also wholly consistent with the recent results of Best et al.(2005)
derived from their SDSS sample of radio-emitting galaxies.

The structure of this paper is as follows: in section \ref{sec:opt} we briefly summarize the properties of the redshift survey used in this work, while in section \ref{sec:multi} we present our multiwavelength database  built for the 15R sample. This database gathers information on $H\alpha$, R-band, radio, far-infrared, and X-ray emission as well as radio and optical morphologies. In section \ref{sec:radio} we compare the classification of galaxies based on standard spectral classification with a classification system based on the radio
and optical morphologies. We outline a classification scheme that merges the information from the two approaches and select clean samples of star-forming galaxies and AGN with radio emission. In section  \ref{sec:fraction} we evaluate the dependence on the environment of star-formation and AGN activity for the 15R sample, highlighting differences 
with similar analyses based on optically selected samples. Our conclusion are summarized in section \ref{sec:conclusion}

\section{The optical sample}
\label{sec:opt}

The 15R-North galaxy redshift survey is a uniform spectroscopic survey (with signal/noise ratio of order 10) covering the range 3650-7400\AA~for 3149 galaxies with a median redshift 0.05 within two $ 2.5^{\circ}$ strips covering a portion of the sky delimited by $8^h \le \alpha \le 17^h$ in right ascension and $26.5^\circ \le \delta \le 29.0^\circ$ or $30.0^\circ \le \delta \le 32.5^\circ$ in declination (B1950). For this survey, 2395 galaxies constitute a  magnitude limited sample 90\% complete to a Kron-Cousins R magnitude $R=15.4$.

 The median slit covering fraction is 24\% of the galaxy, apparently sufficient to minimize the effects of aperture bias on the  $H\alpha$ equivalent widths \cite[]{Carter01}. The spectral types of the 15R galaxies have been classified using ratios of strong emission lines (H$\alpha$, [N II] 6583, [O III] 5007, and H$\beta$), according to the prescription of BPT and Veilleux \& Osterbrock (1987). The spectra have been divided into HII-like (20 \% of the sample), AGN-like (17\% of the sample) and absorption-line spectra (51\%). The remaining 
12\% of the spectra show unclassifiable emission lines and may include a
hybrid population of galaxies with both star formation and AGN activity \cite[]{Carter01}.

 This survey has been analyzed previously to evaluate the environmental dependence of star-formation and AGN activity \cite[]{Carter01}. The results  found with this survey have been confirmed by other authors in larger samples such as 2DF  and the SDSS DR1 (Miller et al. 2003; Mateus \& Sodre$^{\prime}$ 2004). We therefore can compare our results on the environmental dependence of AGN and star-formation activity in radio-emitting galaxies with those found using optical spectral classification alone.

\section{The multiwavelength analysis of the 15R sample}
 
Several biases can affect the standard BTP spectral classification scheme, including aperture bias, dilution of the spectral lines, and dust 
obscuration. For example, it has been shown \cite[]{Moran02} that when the angular size of a Seyfert galaxy is comparable to the slit width, the nuclear 
spectral features can be overwhelmed by the host galaxy light, leading to the exclusion of these galaxies from the sample of AGN. 

Martini et al. (2002) showed that in the cluster A2104, five out of six X-ray emitting AGN do not show any typical AGN optical spectral features and would therefore  
be missed by using spectroscopy as a tool to select AGN. We show here that
spectroscopically unremarkable AGN are a substantial fraction of radio-emitting AGN and, as a result, standard methods of classification for AGN may lead to
substantial incompleteness, while for star-forming galaxies there is general
agreement among classification techniques.

\label{sec:multi}
\subsection{A multiwavelength database}

We cross-correlated the 15R sample with the FIRST 
\cite[]{Becker95}, and NVSS radio catalogs \cite[]{Condon93}, the IRAS Point Source (IPAC 1986) and Faint Source catalogs \cite[]{Moshir90}, and the ROSAT Bright Source \cite[]{Voges99}, Faint Source \cite[]{Voges00}, and WGACAT catalogs \cite[]{White00}.

NVSS and FIRST both sample the radio continuum emission at 1.4 GHz.
We choose to use both NVSS and FIRST in order to gather 
information about the radio morphology of our sources, which is better sampled in FIRST thanks to its higher angular resolution ($5^{\prime\prime}$), and to obtain the best estimate for the flux densities which are more accurate in
the lower-resolution NVSS. Whenever available, NVSS radio flux densities have 
been used. When radio sources were split into several components, we summed
the flux densities from the individual FIRST sources unless there was an unresolved source in NVSS comprised of all components in which case
we used the flux density of the unresolved source as the best estimate.
We measured source sizes using the semimajor axis from an elliptical Gaussian fit to the source surface brightness distribution or, in the
case of multiple-component sources, measured extents directly from the images.

IRAS sources are mostly not resolved, making impossible a comparison with FIRST for the region of the emission within the sampled galaxies.
We selected only point-like ROSAT sources. Extended X-ray emission associated with galaxies might arise in a hot intracluster medium surrounding the galaxies which is not directly relevant to galaxy activity.

In order to define the optimal search radius in all the cross correlations, we built for each catalog four false catalogs by offsetting all positions by
$1^{\prime}$ in each of the four cardinal directions. We calculated the number of false matches expected for different search radii by cross-correlating these false catalogs with the 15R catalog. 

The final search radius $R_s$ was chosen to strike a balance between having the largest number of real matches and a modest number of spurius ones. Generally speaking we tended to use larger radii and to check by hand the more
distant matches, even when the nominal positional accuracy of the sources was
high. When more then one source was associated with an optical galaxy, we checked those matches and assigned the source using the images available.
This avoided missing the large extended radio sources which are resolved into multiple radio components by FIRST; it also included sources for which the peak of the radio emission was not centered on the optical galaxy centroid which defines the  center of our search radius. Selecting the radio sample by means of
an automated procedure could miss these features and lead to incompleteness.

The search radii used were $R_s=18''$ for the FIRST survey, $R_s=28''$ for NVSS, $R_s=7''$ for 2MASS, and $R_s=60''$for ROSAT. Based on our matching to false
catalogs, these radii were selected to ensure 90\% completeness and 90\% accuracy in the final catalog. In the case of the IRAS catalogs, we took into account the ellipticity of the position error regions, using ellipses instead of circles to find the IRAS-galaxy matches. The search ellipses vary for each source and the major and minor axis are a multiple of the major and minor axes of the error 
ellipses given in the IRAS catalog. We allowed these ellipses to be stretched by different amounts and checked the false matches expected in each case. We chose 5 as best stretch factor. For the magnitude-limited sample of 15R we find 315 matches with FIRST, 370 matches with NVSS, 312 matches with IRAS, 108 with 
ROSAT.


The final radio sample is comprised of 520 sources; 79 of them have emission detected only in FIRST, while 166 are detected only in NVSS. FIRST has a lower detection threshold than NVSS for compact sources (1.0~ vs 2.5~mJy) 
accounting for the FIRST-only detections; on the other hand, FIRST has much higher resolution and tends to miss some low-surface brightness extended radio sources detected in NVSS. The systems detected in only one of the two radio surveys constitute a sizable fraction of the total radio sample (46\%); use of both surveys is essential for defining a large radio sample with a flux density limit of $\sim 1-2$~mJy.

\subsection{An independent classification of radio sources}

Radio emitting galaxies are either star-forming, host an AGN, or are a combination of the two.
Star-forming radio galaxies are typically late-type galaxies and their radio emission is usually aligned with the galaxy's optical emission, or centered on it for face-on objects.
Radio AGN are either point-sources in late type galaxies or radio sources in early type galaxies. Jets are generally not aligned with the optical emission,
being frequently found perpendicular to the galaxy's major axis. Classification of the type of activity in a sample of radio-emitting galaxies can be done successfully if one has a reliable classification of the morphology of the radio emission and either the host galaxy type (early-type or late-type) or, at least,  the relative alignment of the radio and optical emission. We undertook
to classify the morphologies of our radio sources and their host galaxies using
this information. 

For each radio source, two cut-outs of the radio images were produced from the 
cut-out server of the FIRST catalog (White et al. 1997): $5'\times 5'$ and $1'\times 1'$.
For the radio classification, we examined all cut-outs associated with galaxies in 15R and used the major axis of the radio sources given in the FIRST catalog as an additional indicator to discriminate between extended and point sources (point sources are defined as having a deconvolved major axis smaller than $2.5^{\prime\prime}$). This led to the division of radio sources into five broad categories: FRI (bright nuclear point source and faint radio lobes -- Fanaroff \& Riley 1974), FRII (bright radio lobes, faint nuclear source), extended sources, point 
sources, and sources for which the morphology was unclear. 

In order to explore further the nature of the radio sources with ambiguous
morphologies we used the optical morphological classification of their hosts.
The slice of the universe covered by the 15R survey has been widely studied by several authors and many galaxies in this survey have known morphologies.
We therefore searched for the optical morphologies for 15R galaxies using the  NED database. We were primarily interested in simply dividing our sample of galaxies into
early types and late types, since the two classes differ substantially in terms of ongoing star-formation activity. We were able to find published classifications for  49\% of the hosts of our radio sources.  Most of these galaxies are listed in the Third Reference Catalogue of Bright Galaxies \cite{dev95}. In additon, we established the optical morphology for a further 10\%  of the radio sample by looking at the DSS images. The large majority of this small fraction of galaxies classified by eye are nearby galaxies that show clear spiral structure.

In the final sample of 260 galaxies with radio emission and available optical classifications,  29\% are classified as early-type (ellipticals or S0) and the rest as late type. The ratio of early-type to late-type galaxies for our final sample of classified galaxies, 0.41, is in good agreement with the fraction of early-type to late-type objects found in other surveys with systematic classification (e.g., the Century Survey \cite[]{Geller97}). 

These optical morphologies have been used together with the radio morphology to obtain our classification of active galaxies: extended radio sources associated with late-type galaxies are classified as star-forming, point-like radio sources associated with late-type galaxies are classified as Seyferts (AGN), extended radio sources associated with ellipticals are classified as jet-like AGN and point-like radio sources (or unresolved radio sources from NVSS) associated with ellipticals were also classified as AGN. 

For extended radio sources with no available host optical classification, we took advantage of the image of the galaxies on the DSS plates and compared the alignment of the radio and optical emission.  If the radio source was extended and the radio emission was aligned with and/or centered on it we classified it as a star-forming galaxy. If the radio source showed no alignment with the major axis of the optical emission, we classified it as a jet-type AGN.

One potential problem with this classification is that it might bias us towards excluding nuclear starbursts, since spiral galaxies with point-like radio  emission are all classified as AGN (Seyferts); however we show 
in section \ref{sec:radio} that our morphological classification for these galaxies largely agrees with the spectral classification and thus this should 
not be a significant problem. 
Furthermore, the implied fraction of galaxies classified as radio emitting Seyferts in this bright optical sample is roughly consistent with what has been found in previous work.
We have 2590 galaxies in 15R with $R<15.4$; the vast majority 
have absolute magnitudes $-18>R>-22$. We have classified 106 point-like radio sources as AGN which gives a fraction of $\sim$4\%.
 For comparison  Huchra and Burg (1992) find in the CfA Survey  a fraction of 1\% and Ulvestad and Ho (2001)  find a fraction of $9.6\pm 3\%$ in the Palomar spectroscopic survey. 
We also checked the optical images for the FR1 and FR2 sources to make sure that the classification was correct and there were no optical galaxies associated with the putative lobes. 
Sources with no clear classification have been classified as either extended (EXT) or point-like (PNT) if detected in FIRST and left unclassified (UNC) if detected only in the NVSS.

\begin{figure}
\includegraphics[scale=0.7]{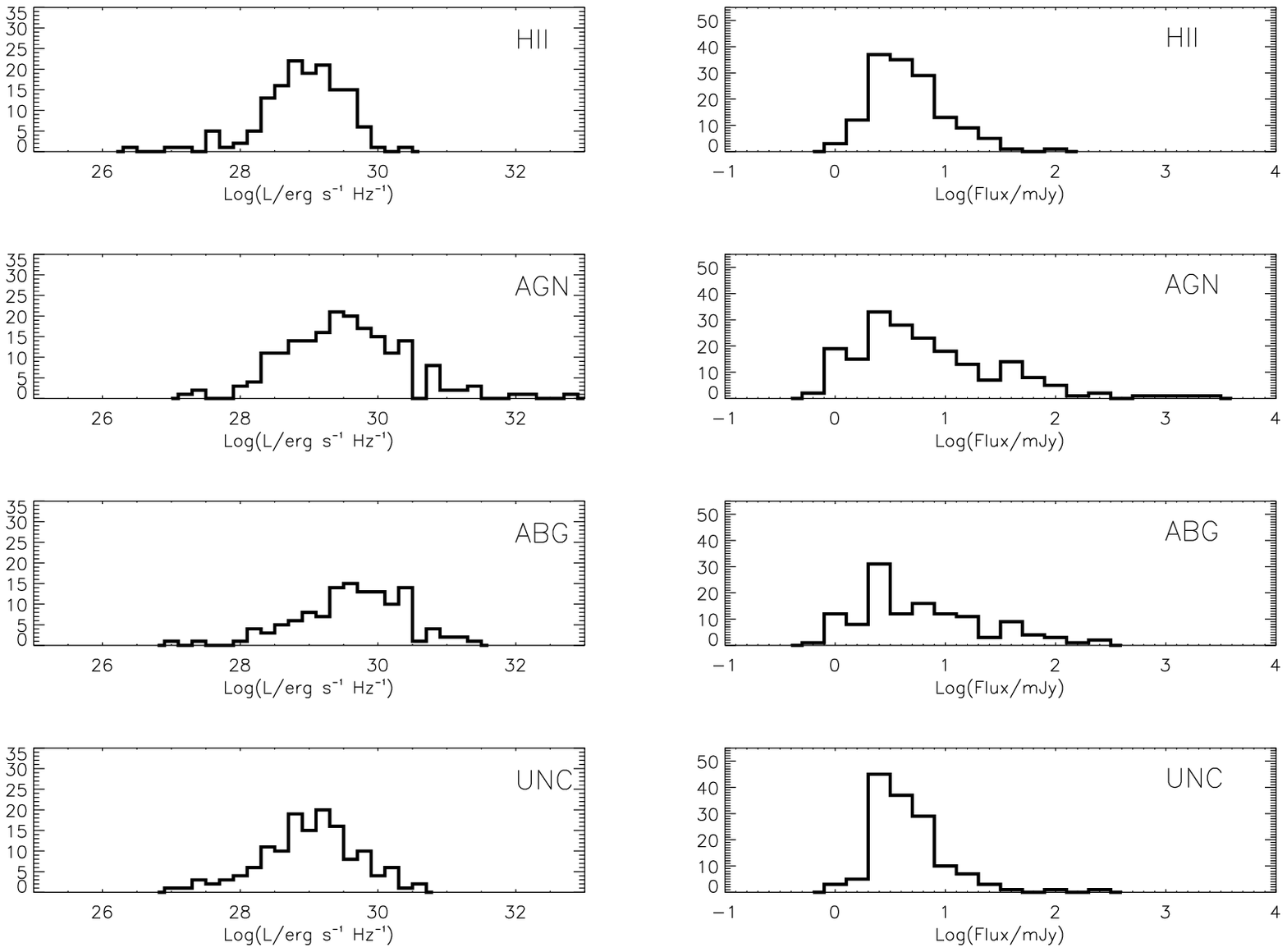}
\caption{\emph{Distributions in luminosity and flux densities for the classified radio sources}}
\label{radio_hist}
\end{figure}

The distribution in radio luminosities for different classes of objects is shown in Fig. \ref{radio_hist}

\section{Comparison of the Two Classification Schemes}
\label{sec:radio}

We first compared the classification system based on the standard BTP diagram and our classification based on radio (and optical) morphology for these radio-emitting galaxies.
Radio-emitting galaxies must be either star-forming or have AGN activity (or a combination of both). How do the two classification schemes perform in classifying this subsample of active galaxies?

Of the 520 galaxies in 15R detected in either NVSS or FIRST, we find
138 AGN, 113 HII, 67 PNT, 52 EXT, and 150 UNC.
The classification of these same sources according to their optical spectra is as
follows: 69 AGN, 155 HII galaxies, 171 weak or strong emission line galaxies
that are not classifiable based on their optical spectra, and 125 absorption-line galaxies (ABG).
Our radio classification system provides a classification for 48\% of the sources, while optically only 43\% have a definitive class assigned.

For emission line systems the two classifications largely agree: 38\% of the spectra-classified AGN are classified as AGN in the radio, 34\% are left unclassified (as either PNT, EXT, or UNC) and 18 galaxies have conflicting
calssifications (we classify them as HII galaxies on the basis of being
spirals with extended radio emsission roughly aligned with the optical morphology. Likewise, 33\% of the spectra-classified HII galaxies are classified by radio morphology as HII, 53\% are left unclassified, and 15 galaxies are
classfied as AGN on the basis of their radio morphology.

The large fraction of sources which are classified with one method but not the other suggests that using these two method of classification together will significantly enhance the number of galaxies with reliable classifications. We describe a joint approach in section \ref{sec:finalsample}

A total of 38\% of the galaxies we classify as Seyferts (late type galaxies with a point-like radio source) are classified as AGN according to their spectra, with only 14\% classified as HII galaxies (the rest have unclassified spectral types). This shows substantial agreement between the two classification schemes for this class of objects. The 14\% of galaxies with conflicting classifications may be nuclear starbursts or composite objects misclassified in our scheme as pure
Seyfert galaxies. We partially rectify this by using the ratio of FIR to radio 
emission, as discussed in \S6.

The most interesting result of this comparison is the fact that while the fraction of HII galaxies classified are similar with the two methods, the fraction of AGN is much different: with the radio classification scheme, 27\% of
the radio sources are classified as AGN, while with spectral classification only 13\% of the sample is so classified.

To understand the origin of this discrepancy we examined the spectral classifications of our radio AGNs: only
19\% of them are classified as AGN optically, while 28\% are listed as unclassifiable emission line systems and 15\% as HII; the plurality --  38\%  -- are classified as ABG.
This means that more than one-third of our radio AGN are in ABG systems and 
are therefore missed by the standard BTP classification scheme.
Since the ABG systems represent 24\% of our whole sample of radio sources, the radio AGN among this group represent $\sim 10\%$ of our whole sample, a fraction comparable to the 13\% contributed by emission-line-classified AGN; i.e.,
the AGN we identify are not recognizable as AGN from their optical spectra.

This class of spectroscopically unremarkable objects have distinct morphologies: 52\% of these radio sources are classified as FRI, FRII, jet, or simply extended, while this percentage drops to 20\% in radio AGN recognized as such from their optical spectra. 
For those with known optical morphologies, we find that 82\% reside in ellipticals or lenticular galaxies, while emission-line AGN, on the contrary, are found preferentially in late-type galaxies (62\% of radio AGN with emission-line spectra are associated with spirals or irregulars) and tend to be radio point sources (77\%). Roughly 10\% of the absorption-line AGN show associated X-ray emission. A similar fraction of X-ray emitting sources is found
among the emission-line AGN.

As noted earlier, we independently classified our sources based only on their X-ray emission: galaxies with X-ray luminosities higher than $10^{42}$ erg s$^{-1}$ are classified as AGN, since no pure starburst galaxies have been identified with higher luminosities (Ranalli et al. 2003); there are, however,
AGN below this threshold, but this cut provides a clean sample of AGN which is of principal interest here. In this sample of 67 X-ray-selected AGN, with or without radio emission, 49\% have absorption-line spectra and 51\% emission-line spectra. The population found by Martini et al. \cite[]{Martini02} are representatives of
the former sample.

The nature of the difference between absorption-line AGN and emission-line AGN is unclear. It is possible that this can result from an effect of dilution of the spectral lines caused by the light of the host galaxy, although a large fraction of this kind of system is also found in 2dF and SDSS where better spectra are available (Reviglio 2003; Best 2004; Best et al. 2005). More importantly, these objects represent a class with radio properties which are rare in the emission-line systems, suggesting a physical link between the morphology of the optical galaxy, its radio AGN morphology, and the lack of emission-lines.
Kauffman showed that AGN with strong OIII emission preferentially inhabit massive galaxies with young stellar populations, suggesting the possibility that the existence of these AGN depend on the gas content of the host galaxy.

X-ray emitting AGN do not seem to favor either of the two spectral categories. If X-ray emission is to be regarded as the principal signature of the process of accretion onto supermassive black holes that powers AGNs, then the fact that X-ray emission can be found in galaxies of both spectral types suggests that spectral lines may be just a signature of the galactic environment in which the AGN happens to reside (gas-rich or gas-depleted) and therefore of the morphology of the host. A  reservoir of cold gas in the host galaxies, as traced by the presence of emission-lines, does not seem to be required for powering an AGN, since radio- and X-ray-emitting AGN are clearly present in line-free systems.
This is similar to the case for radio quasars where the radio-loudness is unrelated to the amount of gas present in the host galaxy \cite[]{Dunlop03}.

Another possibility is that these systems are heavily obscured. However, this would pose the question of why high obscuration should be at work in ellipticals and not in spirals, where absorption-line AGN systems are rarely found. Furthermore, if strong dust obscuration is at work in systems lacking emission lines, one would expect a higher fraction of radio AGN with absorption lines to have FIR emission from hot dust reradiation than emission-line systems. On the contrary we find that only 3\% of the absorption-line radio AGN have detected FIR emission at $60 \mu$m, while the emission-line systems have a 
far-IR-detected fraction of 25\%.

\subsection{Below the faint limit}
For any study examining the properties of galaxies, it is essential to know
if the sample of galaxies selected is representative of the underlying
population. We have shown that emission-line subsamples exclude a large
fraction of AGN and thus introduce a potential bias. It is appropriate to ask
whether our radio-selection introduces a similar bias, since it is clear that
our subsample is much smaller than the emission-line fraction of the
15R Survey; i.e., are galaxies with emission-line or absorption-line spectra
that are undetected in our radio surveys physically different, or is their 
radio emission just too faint to be detected?

While a definitive answer must await the next generation of radio surveys, we
have taken a first step toward understanding the properties of galaxies below the limit of the current surveys by employing a stacking procedure developed by Glikman et al. (2004) which derives the mean radio flux density for any
class of objects with undetected radio emission in the FIRST Survey.

Applying this procedure to all 1078 absorption-line galaxies in 15R with no radio emission detected, we obtain a $3\sigma$ detection of a point-like
radio source with a flux density of 0.02~mJy.
A similar analysis of the 74 undetected, optically classified AGN yields
a mean flux density of $0.41\pm0.04$~mJy, while the 197 stacked HII galaxies
reveal, as expected, an extended source with a lower mean flux density of $0.19\pm0.04$~mJy.

It is worth noting that if a galaxy is star-forming, it must have radio emission.
The exceptions might be objects that have lost their population of cosmic rays
which are not being resupplied by ongoing supernovae, such as might occur at the end of
a starburst; however, this phase will be short-lived, and such instances
should be rare. It is less clear if all galaxies with AGN activity must also have radio emission. The radio-quiet/radio-loud AGN dichotomy has been a matter
of debate for some time (cf. Cirasuolo et al. 2003 and Ivezic et al. 2004 for
a recent debate on this issue). While the precise shape of the radio luminosity
distribution and the presence in the sample of truly radio-silent objects
cannot be determined from the stacking procedure, it is clear that the mean
radio flux density of undetected AGN candidates is only a factor of $\sim 2$
below the FIRST survey threshold.

\section{A merged classification scheme}
\label{sec:finalsample}
We classify our sample of 520 galaxies by integrating the information from the different classification schemes. 
We classify all sources as HII, AGN or unclear (UNC).
When the radio and spectral classifications agree, we classify the source as either AGN or HII. When only one scheme provides a definite classification (as either AGN or HII) and the other has no classification, we adopt the classification
available . When the two classifications disagree, or in the cases for which
neither scheme suggests a source type, we classify the source as unclear. We also added a further AGN indicator: galaxies with X-ray emission stronger than $10^{42}$ erg s$^{-1}$ have been classified as AGN, since only active galactic nuclei can power such strong X-ray emission; these represent a small fraction of our sample. The result of this merging is as follows: 212 AGN, 186 HII, and 122 
UNC. 

We note that now 77\%  of our sample has a classification, a much larger fraction than the 48\% obtained solely with the radio classification and the 43\% obtained with the spectral classification alone. Half of the UNC objects  (12\% of the total radio sample) result from conflicting classifications; many
of these might be hybrid systems exhibiting both star-formation and AGN 
activity.

\section{Further classifications based on photometry}

\label{sec:AGN}
In order reduce further the number of objects with the UNC classification, we examined
two methods developed in the past by Machalski and Condon (1999) which rely
solely on radio, FIR and optical photometry. According to these authors, the 
separation of star-forming from AGN galaxies can be made by calculating one of two parameters, $R$ and $Q$. $R$ represents the radio-to-optical ratio defined as $R=log(S_{1.4}/{f_R})$ where $S_{1.4}$ is the radio flux density at 1.4 GHz expressed in mJy and $f_R$ is the flux density at 0.70 microns (photometric R-band) given by  $f_R=2.87\times 10^{6-0.4R}~{\rm mJy}$. Radio loud objects are defined as having higher $R$ parameters. $Q$ compares the total far-infrared emission to the radio emission and is defined (Helou et al. 1985) as $Q=log(\frac{FIR}{3.75\times 10^{12} Hz)}-log(S_{1.4}\times 10^{-29})$ where $FIR=1.26\times 10^{-17}(2.58S_{60\mu m}+S_{100\mu m})~{\rm mJy}$. Radio-loud objects tend to have lower $Q$ parameters.
According to the R-parameter criterion AGN have $R>1$ while star-forming galaxies have $R<1$; likewise, $Q>1.8$ indicates a star-forming galaxy, while $Q<1.7$ indicates an AGN-dominated source.

\begin{figure}
\includegraphics[scale=0.7]{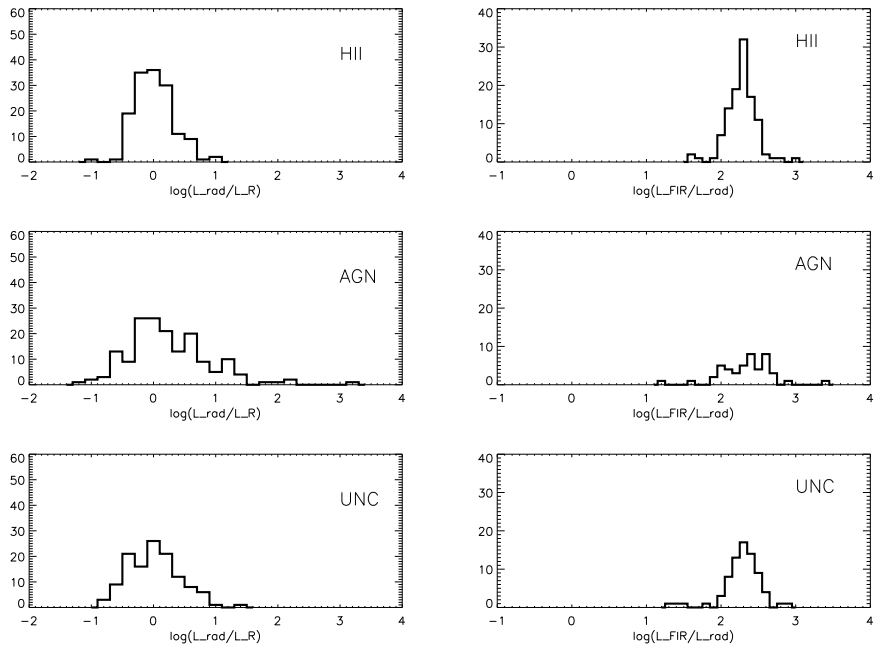}
\caption{\emph{Distribution of the R parameter (first column) and Q parameter (second column) for the classified radio sources}}
\label{QR_par}
\end{figure}

We have assessed the completeness and accuracy of these criteria for the selection of AGN by
applying them to our sample of classified galaxies.
Figure \ref{QR_par} shows the distribution of the $R$ parameter for the three classes of radio sources: HII, AGN and UNC.
From these plots it is clear that AGN and HII galaxies do not divide cleanly into two classes with different $R$ parameters, largely as a consequence of the many AGN classified as point-like radio sources in spiral galaxies. For these sources the radio emission is fainter than for the extended AGN such as
FRI or FRII galaxies, and is comparable to that of star-forming galaxies. 
The distribution of the $R$ parameter for these low-luminosity AGN and star-forming galaxies are similar.

The radio-to-FIR emission ratio given by the Q parameter, yields the distribution presented in figure \ref{QR_par}. The populations of low-luminosity AGN and star-forming galaxies do not differ significantly in this case either. However, we note that the HII galaxies decidedly outnumber the AGN: most radio AGN lack strong FIR emission. Therefore, in order to classify additional galaxies in the unclear class, we have adopted the FIR emission as another indicator of ongoing star-formation. The contamination of the FIR-emitting AGN we introduce is only $\sim$ 20\% of the UNC galaxies with FIR emission or about a dozen objects
(or $\sim 7\%$ of the total HII galaxies).

\begin{figure}
\includegraphics[scale=0.7]{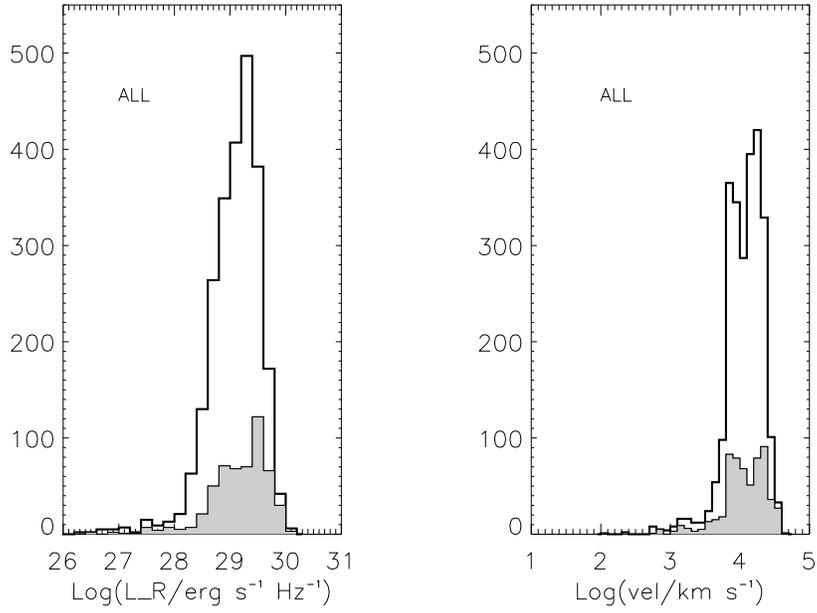}
\caption{\emph{The optical luminosity and redshift distributions for the 15R
galaxy sample (unshaded histogram) compared with the radio-detcted portion of
the sample (shaded)}}
\label{fullhist}
\end{figure}

\begin{figure}
\includegraphics[scale=0.7]{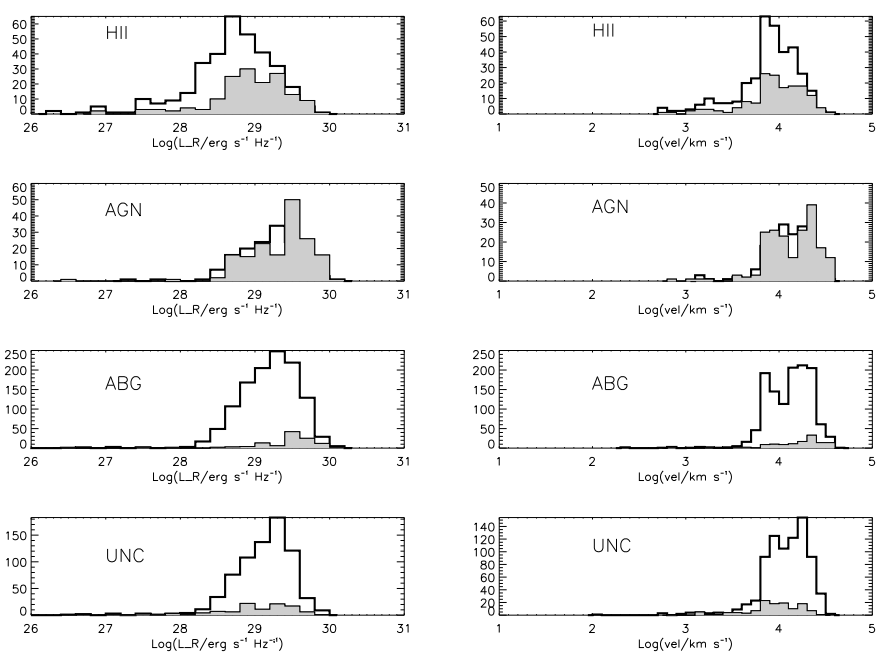}
\caption{\emph{The optical luminosity and redshift distributions for the various classes of galaxies. Open histograms represent the optical classification, while the filled histograms show the radio classification. The radio ABG galaxies,
ultimately classified as AGN in our scheme, show the same distribution as the rest of the radio AGN sample.}}
\label{class_hists}
\end{figure}

We obtain a final sample of radio-emitting galaxies for the analysis of the environmental dependence of the star-formation and AGN activity which has 223 HII, 212 AGN and 85 unclassified galaxies. In figure \ref{fullhist} we display the optical luminosity ($L_R$) and redshift distributions for the 15R survey as 
a whole and compare it to the radio-detected subsamples. Unsurprisingly, the 
optical luminosities of the radio sources are biased high compared to the distribution of whole sample, reflecting the general correlation of radio and optical luminosities and the flux-limited nature of the radio survey. The redshift distributions are similar; the larger fraction
of detections at the highest redshifts reflect the small number of radio-loud
AGN at the highest luminosities. In figure \ref{class_hists}, we show the
same distributions for the galaxies by class, distinguishing between the optical
(unshaded) and radio (shaded) classification schemes. The radio-emitting ABG
galaxies, concentrated at the highest optical luminosities, are ultimately included in our radio AGN sample as discussed above.

\section{Environmental Dependence of Radio-emitting AGN and Star-forming Galaxies}
\label{sec:fraction}

Previous studies have claimed that the fraction of emission-line AGN remains 
constant across several orders of magnitude in local galaxy density \cite[]{Carter01, Miller03} or even decreases with increasing environmental density. We have shown here the existence of a large fraction of radio-emitting AGN which lack emission lines, all of which are missed in work based on
optical spectral classification. Since these sources are most frequently associated with early type galaxies, we expect them to populate denser environments. 

In order to investigate the correlation of star-formation and AGN activity with the density of the environment, we use a magnitude-limited sample with galaxy redshifts in the range $0.0033<z<0.075$ in order to avoid the Virgo infall region where the measure of the redshifts are unreliable and to avoid the distant, poorly sampled regions of the survey. The final sample of classified radio sources in 
this redshift range is 372 sources. 

To associate a volume density with each galaxy, we have implemented the 
procedure outlined in Carter et al. (2001) for magnitude-limited samples. The idea of this method is to calculate the relative densities of different environments by considering the volume enclosing the ten nearest \emph{observed} neighbors of a given galaxy j and dividing by the number of galaxies \emph{actually} present in the volume by the volume itself $V_j$. In order to find the number of galaxies $N_j$ actually present in $V_j$, it is necessary to correct the number of galaxies observed in the volume (ten, by definition) to account for those which fall below the magnitude limit of the sample. Assuming that the Schechter optical luminosity function is indeed universal (changes in the functional form in different environments are not too large -- e.g. Christlein 2000), the correction can be obtained by considering that in order to have one galaxy with absolute magnitude M in a unit volume, there must be a certain number of galaxies, as defined by the luminosity function, which are less bright in the same volume. 

The estimate of the environment density for the $j^{th}$ galaxy we used is therefore:
\begin{equation}
\rho_j=\frac{N_j}{V_j}
\end{equation}

where $N_j=\frac{\int_{-\infty}^{+\infty} \phi(M)d(M)}{\int_-\infty^M_i \phi(M)d(M)}$ with $M_i$ is the faintest absolute magnitude detectable in the survey at the $j^{th}$ position and $\phi(M)$ is the Schechter luminosity function. 
Since the absolute magnitude of galaxies effectively ranges not between plus and minus infinity, but a maximum and minimum value, we have adopted the maximum and minimum absolute magnitude in the survey to calculate the integrals.
For galaxies close to the survey borders, we have corrected the estimate of the density by evaluating the number density of galaxies inside the fraction of the volume given by the intersection of their spheres with the survey borders.
Since 15R photometry has been calibrated with the Century Survey, we used the Century Survey  luminosity function, with Schechter's parameters $\phi_*=0.025 \rm{Mpc^{-3}}$, $\alpha=-1.17$, $M_*=-20.73$ (Geller et al. 1997). Distances
have been calculated from redshifts using the luminosity distance $D_{Mpc}$ for 
a universe described by $\Omega_0=1$ and $H_0=70 \rm{km \ s^{-1}Mpc^{-1}}$.
The R-band absolute magnitudes used  have been K-corrected and also corrected for Galactic absorption.

\begin{figure}
\includegraphics[scale=0.5,angle=90]{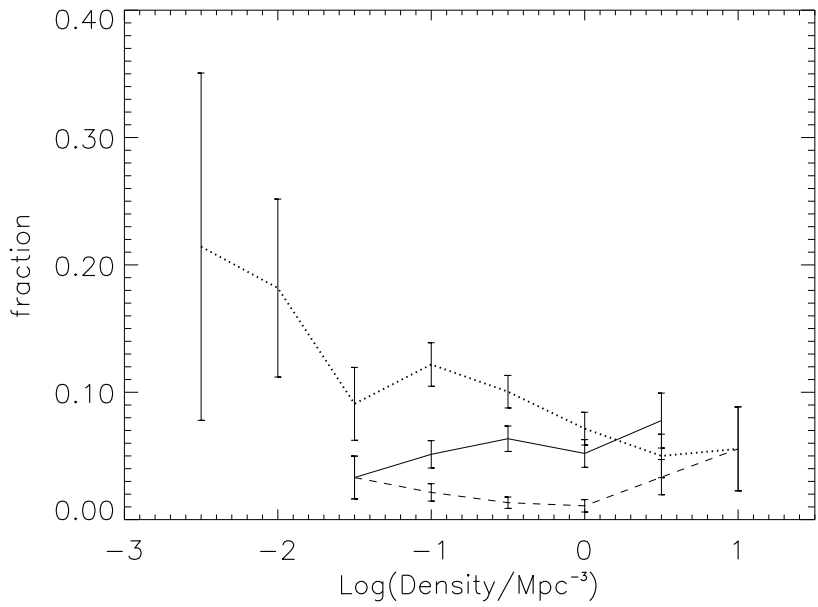}
\caption{\emph{The fraction of galaxies with star-forming activity (dotted line), AGN activity (solid line), and unclassified activity (dashed line).
The decreasing fraction of HII galaxies with increasing density is significant at the 98\% confidence level. The increase of AGN activity with density is only significant at the 82\% level.}}
\label{plots_trends+UNC}
\end{figure}

After obtaining the density surrounding each galaxy, we grouped the galaxies in
logarithmic bins of 0.5 and determined the fraction of galaxies in each bin
with star-forming or AGN activity. The results are displayed in figure \ref{plots_trends+UNC}

The errors are assigned assuming a Poisson distribution; statistical errors are larger than any other source of error. The points in the highest and lowest bins should be regarded with caution since they are calculated with just a few galaxies, although we have excluded from all of plots the points derived from fewer than three galaxies.
Figure \ref{plots_trends+UNC} suggests that the star-forming fraction decreases as the galaxy density increases, while the opposite is true for the AGN fraction which increases as the density increases. 
These two trends are such that the fraction of radio sources mildly decreases with density.

For the star-forming galaxies, the trend shown by our data is consistent with
the results of Carter et al. (2001): the fraction of star-forming galaxies decreases when we move from the field into denser regions such as galaxy clusters. Since star formation is typical of spiral galaxies, and spiral 
galaxies tend to inhabit the low-density regions, this result can be regarded 
as a consequence of the morphology-density relation. These results are also in agreement with those of other authors (e.g., Mateus and Sodre$^{\prime}$ 2003)

For the AGN sample, on the contrary, our results disagree with those found by Carter et al. (2001) using 15R, as well as with more recent studies by Miller et al. (2003) using the Sloan Digital Sky Survey and Mateus and Sodre$^{\prime}$
(2003) using the 2DF Survey. All these studies showed the fraction of AGN to
be independent of local galaxy density. Even more discrepant with our results 
are those by Kauffmann et al. (2004) which find a decreasing fraction of
AGN with increasing galaxy density.

\begin{figure}
\includegraphics[scale=0.5,angle=90]{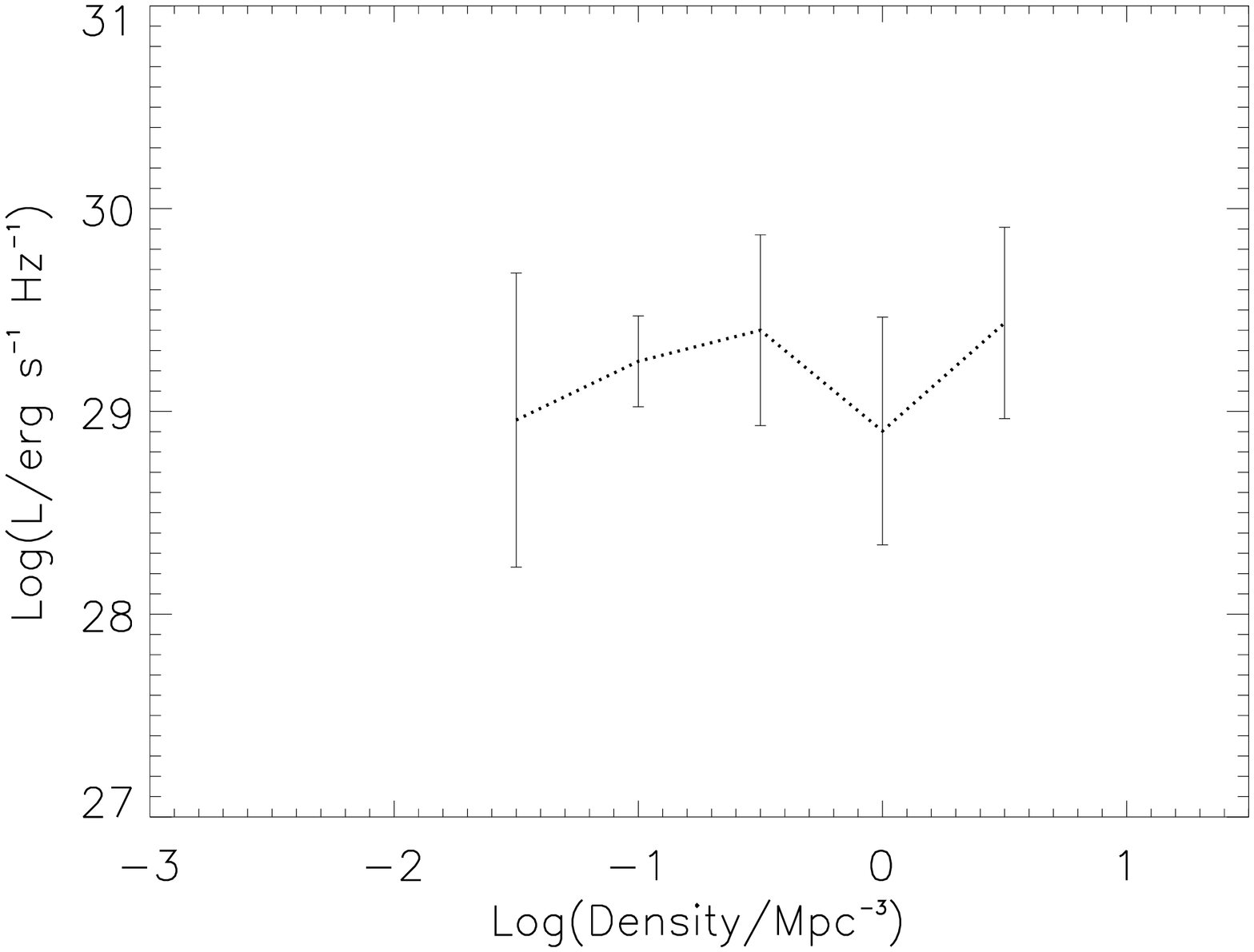}
\caption {\emph{Trend for the median radio luminosity with density for AGN galaxies.}}
\label{trend_rad_lum_AGN}
\end{figure}

Our observed increase in the fractional abundance of radio AGN is not caused by a general enhancement of the radio luminosity in denser regions which raises more sources above the detectability threshold. Indeed, the median radio luminosity in different density bins for the radio AGN does not vary significantly (see Fig. \ref{trend_rad_lum_AGN}). The same statement is true for HII galaxies; since their median radio luminosity does not vary with density, the decreasing fraction of star-forming galaxies is not a result of differences in strength of their radio emission.

\begin{figure}
\includegraphics[scale=0.5,angle=90]{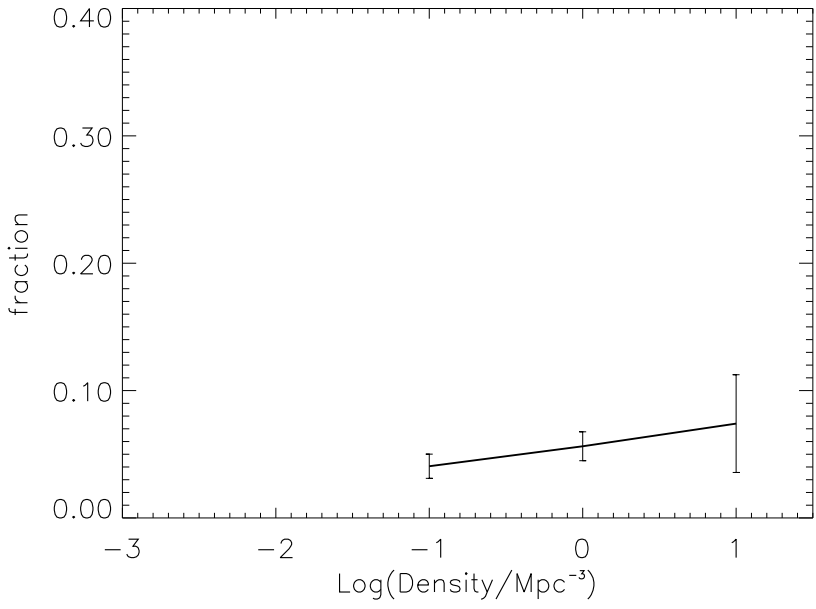}
\caption {\emph{Fraction of absorption-line radio AGN with density}}
\label{final_ABG}
\end{figure}

The difference leading to the different trend for our AGN-classified sources is the inclusion of the large fraction of absorption-line systems with associated radio AGN, which preferentially reside in denser environments. This is shown in fig. \ref{final_ABG}. These sources can be found at all densities, but their fractional abundance is higher in denser ones.  These conclusions, first described in Reviglio (2003), are in agreement with the recent SDSS study of Best et al. (2005).

\section{Conclusion}

\label{sec:conclusion}

By employing a nearby galaxy sample and matching it to the best available radio
surveys (along with surveys in other bands), we have shown that emission-line
galaxy samples are substantially incomplete for identifying galaxies hosting
AGN. Since the absorption line systems hosting AGN are found
preferentially in dense environments, the inclusion of this missed population
of AGN in spectrally defined samples reverses the trend noted earlier: the
total population of AGN {\it in}creases with increasing local galaxy
density. Examination of the stacked radio images for the optical objects
not detected at radio wavelengths shows consistency with the objects in
the radio subsample. It is clear that the use of such properties as radio 
morphology and X-ray emission are valuable adjuncts when classifying objects
from large optical surveys.

\section{Acknowledgments} 

We want to express our gratitude to Margaret Geller for very helpful suggestions in developing this project and for allowing us to use her data prior to publication. We want also to thank Eilat Glikman for enabling us to use her stacking procedure. This work also strongly benefited from discussions with Jacqueline van Gorkom and Antonaldo Diaferio. This work was part of the XVI Cycle of the Dottorato di Ricerca in Fisica
at the University of Torino from which P.R. gratefully acknowledges support.

\clearpage

\clearpage

\clearpage

\clearpage

\clearpage

\clearpage
 
\clearpage
\begin{figure}
\includegraphics[scale=0.5,angle=90]{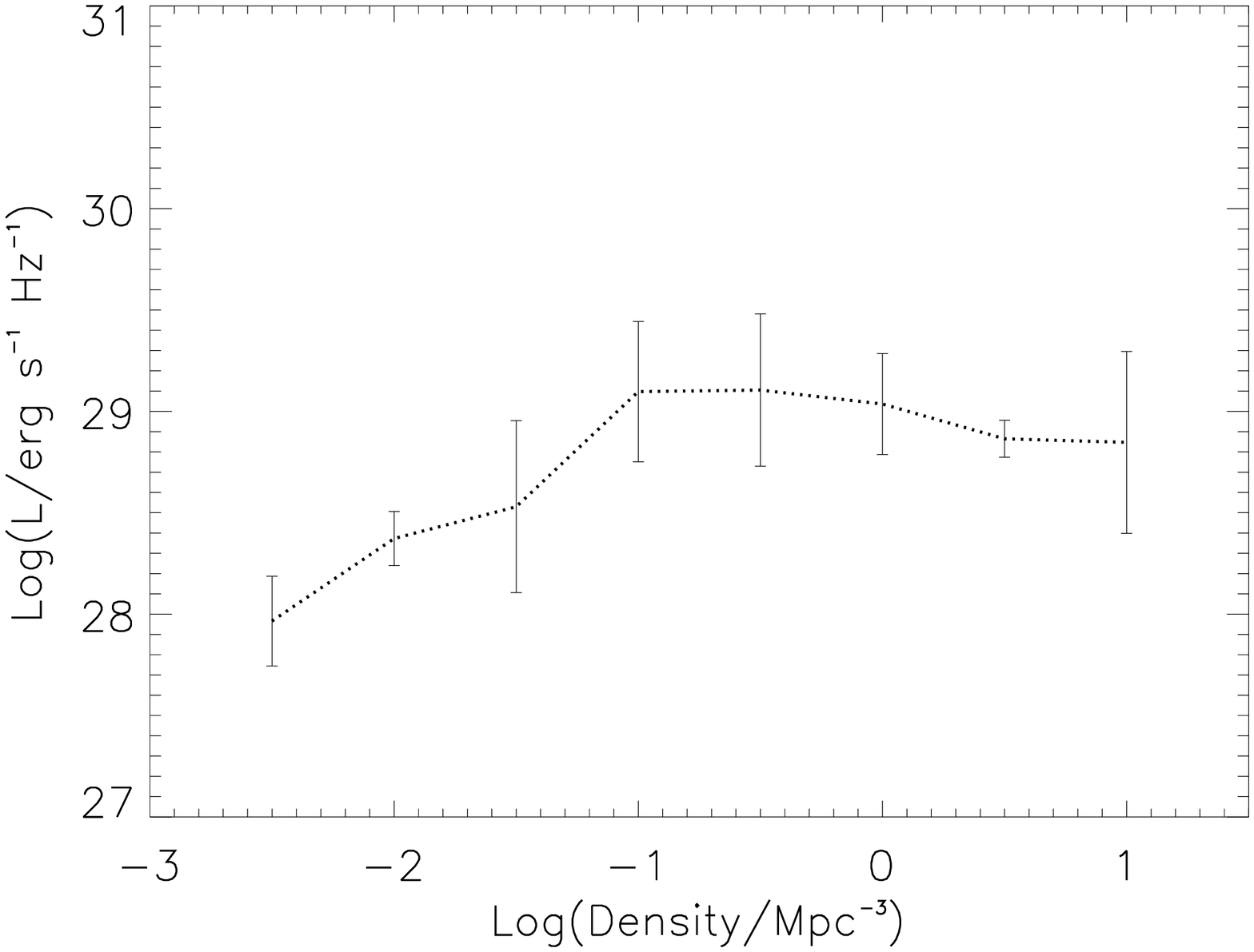}
\caption {\emph{Trend for the median radio luminosity with density for HII galaxies.}}
\label{trend_rad_lum_HII}
\end{figure}

\clearpage

\end{document}